\documentclass[10pt,letterpaper,journal]{IEEEtran}
\usepackage{amsmath,amsfonts}
\usepackage{algorithmic}
\usepackage{algorithm}
\usepackage{array}
\usepackage{booktabs}
\usepackage{pifont}
\usepackage{xcolor}
\usepackage{multirow}
\usepackage[caption=false,font=normalsize,labelfont=sf,textfont=sf]{subfig}
\usepackage{textcomp}
\usepackage{stfloats}
\usepackage{url}
\usepackage{verbatim}
\usepackage{graphicx}
\usepackage{pgfplots}
\usepackage{cite}
\pgfplotsset{compat=1.18}
\usetikzlibrary{plotmarks}
\newcommand{\cmark}{\textcolor{green!50!black}{\ding{51}}}
\newcommand{\xmark}{\textcolor{red}{\ding{55}}}

\begin{document}

\title{SCSM: A Traffic-Native Foundation Model for Transferable\\ Website Fingerprinting}

\author{Xianwen~Deng, Ruijie~Zhao, Mingwei~Zhan, Yijun~Wang, and~Zhi~Xue%
\thanks{Xianwen Deng, Mingwei Zhan, Yijun Wang, and Zhi Xue are with the
School of Electronic Information and Electrical Engineering, Shanghai Jiao
Tong University, Shanghai 200240, China (e-mails: \{2594306528, mw.zhan,
ericwyj, zxue\}@sjtu.edu.cn).}%
\thanks{Ruijie Zhao is with the School of Cyber Science and Engineering,
Southeast University, Nanjing 211189, China (e-mail:
ruijiezhao@seu.edu.cn).}%
\thanks{Corresponding author: Zhi Xue.}}



\maketitle

\begin{abstract}
Website fingerprinting infers the websites visited by users from encrypted
traffic metadata. However, models trained under fixed collection conditions
often degrade as website sets, collection times, network paths, browsers, or
defenses change. Existing transferable attacks either rely on handcrafted
perturbations of individual traces or adapt language-oriented architectures to
traffic, limiting their ability to capture traffic-native semantics. To address
these limitations, we propose SCSM, a traffic-native foundation model for
transferable website fingerprinting. Specifically, SCSM constructs pairs of
pretraining views from the same group of unlabeled traces through Segmentation,
Combination, Scaling, and Masking. These operations produce diverse observable
patterns while preserving the underlying packet events and local traffic
dynamics of real trace fragments. The corresponding windowed traffic counting
matrices serve as inputs for contrastive pretraining of a state-space encoder
without website annotations. The pretrained encoder is then fine-tuned on a
small labeled support set, and predictions are aggregated across temporal scales
at inference. Experimental results demonstrate that SCSM surpasses the
strongest baselines by 16.4\% in average top-3 accuracy over six temporal-drift
tasks on GTT23 and by 13.8\% in average macro-F1 over four cross-domain
datasets. The code and datasets will be made available at
\url{https://github.com/SJTU-dxw/WF-SCSM}.
\end{abstract}

\begin{IEEEkeywords}
Website fingerprinting, traffic foundation models, few-shot adaptation.
\end{IEEEkeywords}

\section{Introduction}
Website Fingerprinting (WF) attacks~\cite{k-FP,CUMUL,AWF,DF,Tik-Tok,Var-CNN,TMWF} infer the websites visited by users from encrypted traffic metadata, such as packet directions, timings, and burst patterns.
Recent deep learning approaches~\cite{ARES,RF,Holmes,CountMamba} have achieved high identification accuracy under controlled experimental settings.
However, their effectiveness largely relies on the assumption that training and deployment traffic follow similar distributions~\cite{juarez2014critical}.
In practice, website content evolves continuously, Tor circuits traverse heterogeneous network paths, users adopt different browser configurations, and the set of monitored websites may change over time~\cite{xie2023rosetta,Palette}.
These factors create substantial mismatches between training and deployment traffic, causing task-specific WF classifiers to deteriorate rapidly~\cite{NetCLR,Swallow,Traverse,Proteus}.
Real-world WF is therefore not a single stationary classification problem, but a family of related recognition problems defined over heterogeneous traffic domains.
A deployment domain may differ from the training domain in its website population, collection period, network location, or browser configuration.
Conventional WF attacks address each domain independently, requiring attackers to repeatedly collect labeled traces and train new classifiers~\cite{NetCLR,Swallow}.
Such a task-by-task learning paradigm fails to exploit the traffic knowledge shared across different WF domains.

Transferable WF attacks~\cite{TF,NetCLR,Swallow,FMWF} seek to alleviate this problem by pretraining traffic encoders and adapting them to a target domain with only a small number of labeled traces.
Existing transferable WF methods typically construct alternative views by
applying handcrafted perturbations to an individual trace~\cite{NetCLR} or its numerical
representation~\cite{Swallow}. This strategy has two limitations.
First, independently modifying burst sizes, packet positions, or window-level
counts can disrupt the directional and temporal dependencies jointly produced
by actual page-loading activities, reducing the fidelity of the augmented
traffic. Second, because every view remains a local variant of a single trace
within a predefined transformation space, the resulting views cover only a
limited range of global traffic structures and browsing compositions.
Consequently, the encoder may primarily learn invariance to designer-specified
perturbations rather than transferable traffic semantics.

More recently, TraVerse~\cite{Traverse} introduces foundation-model pretraining into WF by adapting a pretrained language model to traffic.
It serializes burst sequences as textual prompts, introduces bidirectional attention, and performs traffic-domain pretraining on unlabeled traces through masked traffic-pattern reconstruction.
While demonstrating the potential of large-scale pretraining for transferable WF, TraVerse still inherits tokenization, representation spaces, and architectural priors developed for natural language.
Traffic knowledge is therefore acquired by adapting a language-centric backbone rather than learned natively within a traffic foundation model.

The limitations of both handcrafted traffic augmentation and language-centric pretraining motivate two design requirements for a traffic-native WF foundation model.
First, it should be built on traffic-native representations and directly capture their directional and temporal dependencies, rather than recasting traffic through a language-oriented interface.
Second, its pretraining views should preserve packet events and local traffic
dynamics observed in real traces while introducing diverse higher-level
variations in temporal organization and traffic composition. Such a design
combines event-level fidelity with structural diversity, allowing the encoder
to learn traffic semantics beyond invariance to a fixed set of numerical
perturbations. By satisfying these requirements, the model can learn reusable
traffic representations from unlabeled data and support few-shot adaptation across
website sets, network conditions, and browsing scenarios.

To this end, we propose SCSM, a traffic-native foundation model for transferable WF.
Unlike approaches that serialize traffic through a language-oriented interface, SCSM operates directly on windowed traffic counting matrices, which preserve the temporal evolution of incoming and outgoing packets.
A state-space encoder then models the local dynamics and long-range dependencies within these traffic-native representations~\cite{CountMamba}.
To learn reusable representations from unlabeled traffic, SCSM constructs
traffic-grounded views through three complementary stages. At the trace level,
Segmentation and Combination treat fragments of observed single-tab traces as
construction primitives. They preserve the packet directions and relative
timings within each fragment, while randomized segmentation, cropping,
ordering, and temporal interleaving produce a combinatorial range of global
traffic structures. During matrix construction, temporal Scaling varies the
observation granularity without modifying the underlying packet events. At the
feature level, Masking randomly removes temporal regions, encouraging the
encoder to integrate evidence across the observation horizon.
Applying this pipeline
independently to the same source group yields two structurally diverse views
with the same latent website composition, which form a positive pair for
contrastive pretraining.
The pretrained encoder can subsequently be adapted to new website sets, network conditions, and browsing scenarios using only a small number of labeled traces.

We evaluate SCSM under diverse adaptation settings, including temporal drift,
cross-scenario and cross-domain transfer, open-world recognition, defended
traffic, and multi-tab browsing. Under temporal drift on GTT23, SCSM surpasses
the strongest baseline by 16.4\% in Accuracy@3 averaged over six target weeks.
On Swallow, it improves average
macro-F1 by 6.0\% across
shifts in guard-relay location, website content, and browser. For closed-world
cross-domain transfer, it improves macro-F1 by 13.8\% on average across four
target datasets. Strong results in open-world recognition, defended traffic, and
multi-tab browsing further demonstrate the broad transferability of its learned
traffic representations.

Our main contributions are as follows:
\begin{itemize}
    \item We present SCSM, a traffic-native foundation model for
    transferable WF that operates directly on windowed traffic counting
    matrices. By coupling traffic-native representations with a state-space
    encoder, SCSM learns reusable traffic knowledge from unlabeled traces and
    supports sample-efficient adaptation across website sets, network
    conditions, and browsing scenarios.
    \item We develop a three-stage view construction mechanism comprising
    trace-level Segmentation and Combination, matrix-level temporal Scaling,
    and feature-level Masking. It preserves observed packet events and local
    directional-temporal dynamics while generating structurally diverse views
    through cross-trace composition and multi-level transformations. This
    traffic-grounded design expands pretraining diversity beyond local
    perturbations of individual traces and enables label-free contrastive
    learning from shared website compositions.
    \item We conduct extensive experiments on diverse real-world encrypted
    traffic datasets to assess SCSM's transferability and robustness. Under
    limited labeled supervision, SCSM substantially outperforms state-of-the-art
    baselines, demonstrating the effectiveness and generalizability of its
    learned traffic representations.
\end{itemize}

\section{Related Work}
\subsection{Conventional Website Fingerprinting}
Early WF attacks relied on manually engineered traffic statistics and
conventional classifiers. CUMUL~\cite{CUMUL} represents a trace through its
cumulative packet-size evolution and applies a support vector machine, whereas
$k$-FP~\cite{k-FP} combines traffic statistics with random-forest leaf patterns
to construct website fingerprints. Although effective, these approaches depend
on feature designs tailored to particular traffic characteristics.
Deep learning later enabled WF classifiers to learn features directly from
traffic traces. AWF~\cite{AWF} and DF~\cite{DF} use convolutional networks to
model packet-direction sequences, while Tik-Tok~\cite{Tik-Tok} and
Var-CNN~\cite{Var-CNN} additionally exploit packet timing. TMWF~\cite{TMWF}
and ARES~\cite{ARES} extend WF to multi-tab browsing, and Holmes~\cite{Holmes}
targets identification from incomplete, early-stage traces. More recently,
RF~\cite{RF} and CountMamba~\cite{CountMamba} encode traces as windowed traffic
counting matrices to improve robustness and capture spatial--temporal traffic
structure. Despite their different representations and architectures, these
methods are predominantly trained by supervised learning on a fixed domain.
Their accuracy can therefore decline when website content, network paths,
browser configurations, or monitored website sets differ between training and
deployment~\cite{juarez2014critical}.

\subsection{Transfer Learning for Website Fingerprinting}
Transfer learning reduces the cost of rebuilding a WF classifier for every
deployment domain. TF~\cite{TF} learns an embedding with triplet loss and
performs $N$-shot classification, while FMWF~\cite{FMWF} transfers a
supervised pretrained model to few-shot multi-tab identification. Because
these approaches learn their source representations from labeled traces,
their portability depends on how well the labeled source domains cover the
target traffic conditions.

Subsequent work uses unlabeled traces and traffic-specific augmentations to
learn more general representations. NetCLR~\cite{NetCLR} combines contrastive
pretraining with NetAugment, which perturbs packet bursts to emulate unobserved
network conditions. Swallow~\cite{Swallow} introduces a consistent interaction
feature and loading-time-oriented transformations to align traffic collected
under different network conditions. These methods substantially improve
few-shot transfer, but their learned invariances remain tied to predefined
augmentation operations and parameter ranges.

\subsection{Foundation-Model Pretraining for WF}
TraVerse~\cite{Traverse} introduces foundation-model pretraining into WF by
adapting a pretrained language model to traffic. It serializes burst sequences,
introduces bidirectional attention, and performs traffic-domain pretraining
through masked traffic-pattern reconstruction before training a lightweight
few-shot classifier over the learned representations. This work demonstrates
the potential of large-scale pretraining for transferable WF, but retains
tokenization, representation spaces, and architectural priors developed for
natural language.

\subsection{WF Defenses}
WF defenses seek to obscure website-specific traffic patterns and can be
broadly divided into traffic perturbation and traffic splitting.
Perturbation-based defenses insert dummy packets, delay real packets, or regulate packet
transmissions according to predefined schedules. WTF-PAD~\cite{wtfpad}
adaptively injects padding, while Front~\cite{front} adds dummy packets
according to randomized timing distributions. Tamaraw~\cite{tamaraw} and
RegulaTor~\cite{regulator} regularize traffic transmission, whereas
Palette~\cite{Palette} anonymizes traffic within clusters of websites. In
contrast, TrafficSliver~\cite{trafficsliver} distributes traffic over multiple
Tor circuits so that an observer sees only a partial trace. These defenses
alter the observable packet counts, timings, and trace completeness, requiring
WF attacks to remain effective under defense-induced traffic shifts.

\section{Threat Model and Learning Paradigm}
\label{sec:threat-learning-paradigm}

\begin{figure*}[!t]
  \centering
  \includegraphics[width=\textwidth]{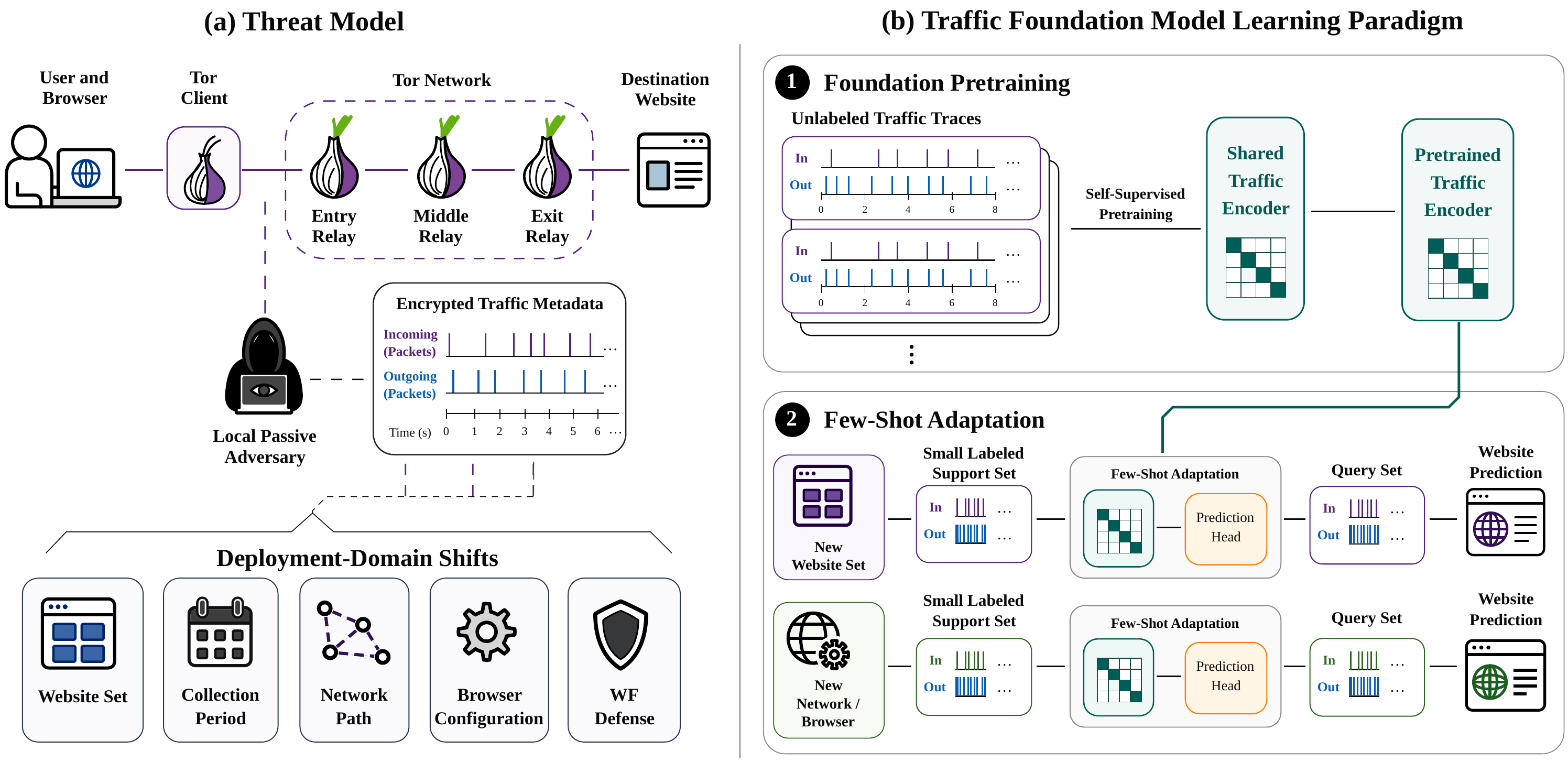}
  \caption{Threat model and learning paradigm of SCSM. (a) A local passive
  adversary observes encrypted packet metadata on the link between the Tor
  client and its entry relay, while deployment domains may differ in website
  set, collection period, network path, browser configuration, or defense.
  (b) A shared traffic encoder is pretrained on unlabeled traces and then
  adapted with a small labeled support set and a target-specific prediction
  head for each downstream task.}
  \label{fig:threat-learning-paradigm}
\end{figure*}

\subsection{Threat Model}
Following the standard WF setting~\cite{juarez2014critical,TF,NetCLR,Swallow},
we consider a local, passive adversary that observes the link between a Tor
client and its entry relay, as illustrated in
Fig.~\ref{fig:threat-learning-paradigm}(a). The adversary records an encrypted
traffic trace $x=((t_1,d_1),\ldots,(t_L,d_L))$, where $t_i$ is the timestamp of
the $i$-th packet and $d_i\in\{-1,+1\}$ denotes its direction, with $+1$ for an
outgoing packet and $-1$ for an incoming packet. The adversary
cannot decrypt packet contents or delay, drop, inject, or otherwise modify the
victim's traffic. Its objective is to infer the website visited by the victim;
in a multi-tab session, the objective extends to identifying the set of
concurrently visited websites.

Unlike the conventional setting in which training and deployment traces are
collected under similar conditions, we allow the deployment domain to differ
in its website population, collection period, network condition, or browser
configuration. The deployment traffic may also
be transformed by a known WF defense. In this case, we assume that the
adversary knows the defense mechanism and can collect a limited number of
labeled traces generated under that defense for adaptation, but cannot
interfere with its operation. We consider both closed- and open-world settings,
as well as single- and multi-tab browsing. These settings require the model to
accommodate changes in both traffic distributions and downstream label spaces.

\subsection{Foundation-Model Learning Paradigm}
We define a \emph{traffic foundation model} as a traffic-native encoder that is
pretrained on unlabeled traces through a task-agnostic objective and can be
reused across downstream WF settings. As shown in
Fig.~\ref{fig:threat-learning-paradigm}(b), it comprises two stages: foundation
pretraining and few-shot adaptation. The first stage learns generalizable
representations from realistic and diverse traffic views rather than
handcrafted numerical perturbations. Let
$\mathcal{U}=\{x_i\}_{i=1}^{N_u}$ denote a corpus of unlabeled traffic traces
used for pretraining.
An encoder $f_{\boldsymbol{\theta}}:\mathcal{X}\rightarrow\mathbb{R}^{d}$ is
optimized with a self-supervised objective:
\begin{equation}
  \boldsymbol{\theta}^{*}
  = \arg\min_{\boldsymbol{\theta}}
  \frac{1}{N_u}\sum_{i=1}^{N_u}
  \mathcal{L}_{\mathrm{pre}}(x_i;\boldsymbol{\theta}),
  \label{eq:foundation-pretraining}
\end{equation}
where $\mathcal{L}_{\mathrm{pre}}$ learns consistency across alternative views
of observed traffic. The resulting encoder is independent of a fixed target
website vocabulary and provides a reusable initialization for downstream WF
tasks.

For a target task $\tau$, let
$\mathcal{S}_{\tau}=\{(x_i,y_i)\}_{i=1}^{n_\tau}$ denote a small labeled
support set and let $\mathcal{Q}_{\tau}$ be a disjoint query set used only for
evaluation. In a $K$-shot single-tab task with class set
$\mathcal{Y}_{\tau}$, the support set contains $K$ labeled traces per monitored
website, so that $n_\tau=K|\mathcal{Y}_{\tau}|$. A target-specific prediction
head $h_{\boldsymbol{\phi}_{\tau}}$ is appended to the pretrained encoder.
Starting from $\boldsymbol{\theta}=\boldsymbol{\theta}^{*}$, the encoder and
head are jointly optimized by minimizing:
\begin{equation}
  \mathcal{L}_{\mathrm{ft}}^{\tau}
  = \frac{1}{|\mathcal{S}_{\tau}|}
  \sum_{(x,y)\in\mathcal{S}_{\tau}}
  \ell_{\mathrm{cls}}\!\left(
  h_{\boldsymbol{\phi}_{\tau}}(f_{\boldsymbol{\theta}}(x)),y\right).
  \label{eq:few-shot-adaptation}
\end{equation}
Here, $\ell_{\mathrm{cls}}$ is the cross-entropy loss. With the pretrained
encoder as initialization, the limited target labels are used primarily to
establish task-specific decision boundaries rather than to learn
discriminative representations from scratch.

\section{SCSM}
\label{sec:scsm}
\subsection{Overview}

\begin{figure*}[!t]
  \centering
  \includegraphics[width=\textwidth]{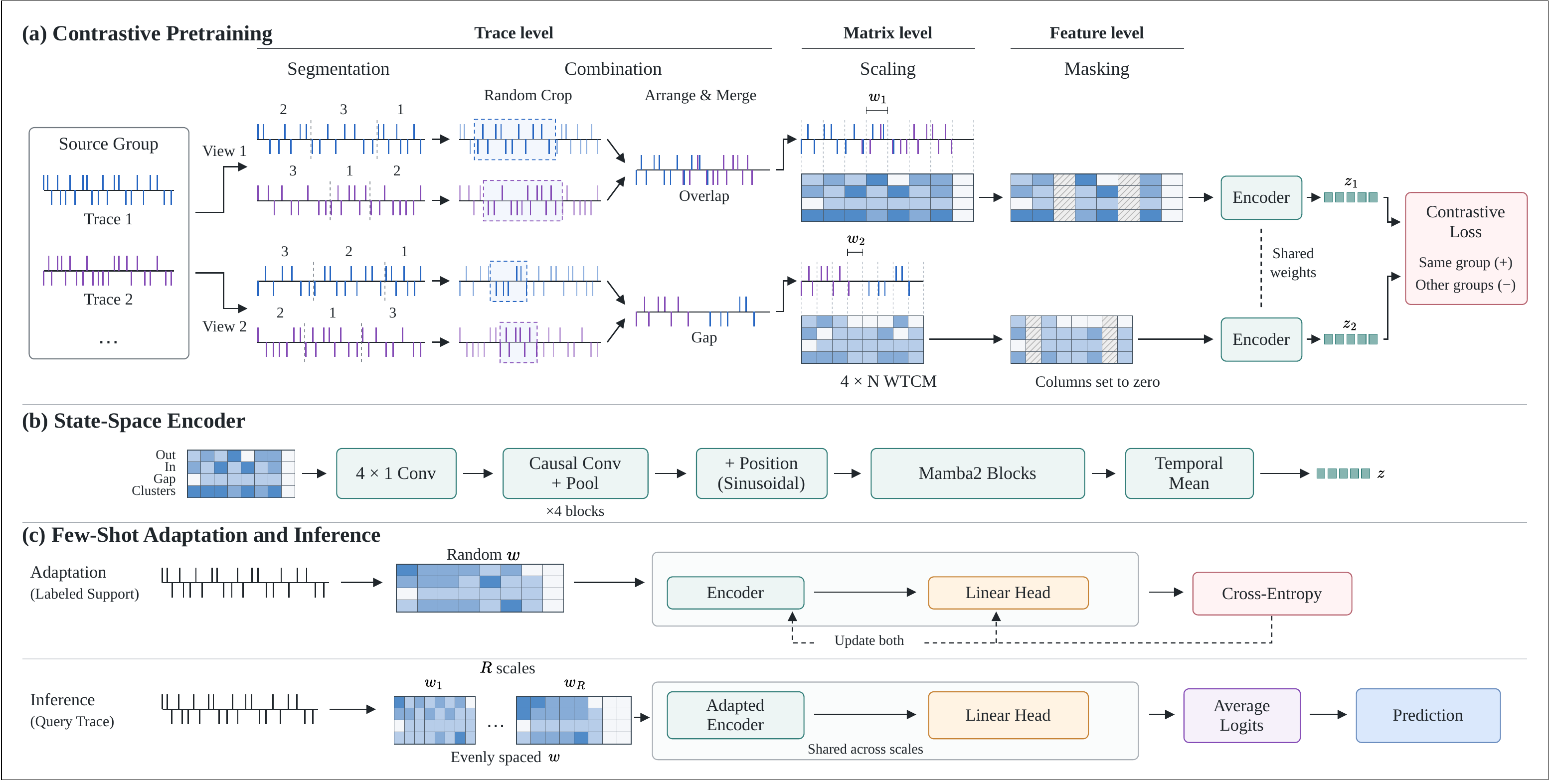}
  \caption{Overview of SCSM. During pretraining, two independent multi-level
  construction branches generate structurally diverse WTCM views from the
  same group of unlabeled traces. A shared state-space encoder learns their
  representations through a cross-view contrastive objective. For a target
  task, the encoder and a linear classifier are adapted on a small labeled
  support set, and class logits are averaged across temporal scales at
  inference.}
  \label{fig:scsm-overview}
\end{figure*}

SCSM learns reusable traffic representations by combining event-level
fidelity with structural diversity. As shown in Fig.~\ref{fig:scsm-overview},
Segmentation and Combination reuse fragments of observed single-tab traces,
preserving their local packet dynamics while varying their global
organization. The composite traces are represented as direction-only
windowed traffic counting matrices (WTCMs), where temporal Scaling and
Masking introduce variations in aggregation granularity and visible regions.
Two independent constructions from the same source group retain a shared
latent website composition, providing structurally diverse positive pairs
without website annotations.

A shared state-space encoder is pretrained to align these paired views
through a cross-view contrastive objective. The resulting encoder is then
adapted with a linear classification head on a small labeled support set,
following Section~\ref{sec:threat-learning-paradigm}. Temporal Scaling extends
to this stage through random window sampling during adaptation and
logit averaging across evenly spaced temporal scales at inference.

\subsection{Traffic-Native Input Representation}
\label{sec:traffic-representation}

\begin{algorithm}[!t]
  \caption{Direction-Only WTCM Construction}
  \label{alg:direction-wtcm}
  \begin{algorithmic}[1]
    \renewcommand{\algorithmicrequire}{\textbf{Input:}}
    \renewcommand{\algorithmicensure}{\textbf{Output:}}
    \REQUIRE Trace $x=((t_i,d_i))_{i=1}^{L}$, window length $w$,
    number of windows $N$, and cluster threshold $\delta$
    \ENSURE Windowed traffic counting matrix $\mathbf{X}(x;w)$
    \STATE Initialize $\mathbf{M}\leftarrow\mathbf{0}_{4\times N}$
    \STATE Initialize timestamp lists $\mathcal{T}_j\leftarrow[\,]$,
    $j=1,\ldots,N$
    \FOR{$i=1$ to $L$}
      \STATE $j\leftarrow\min(\lfloor t_i/w\rfloor+1,N)$
      \IF{$d_i=+1$}
        \STATE $m_{1j}\leftarrow m_{1j}+1$
      \ELSE
        \STATE $m_{2j}\leftarrow m_{2j}+1$
      \ENDIF
      \STATE Append $t_i$ to $\mathcal{T}_j$
    \ENDFOR
    \STATE $j_{\mathrm{prev}}\leftarrow\emptyset$
    \FOR{$j=1$ to $N$}
      \IF{$|\mathcal{T}_j|>0$}
        \IF{$j_{\mathrm{prev}}\neq\emptyset$}
          \STATE $m_{3j}\leftarrow j-j_{\mathrm{prev}}$
        \ENDIF
        \STATE $m_{4j}\leftarrow\operatorname{clusters}_{\delta}(\mathcal{T}_j)$
        \STATE $j_{\mathrm{prev}}\leftarrow j$
      \ENDIF
    \ENDFOR
    \RETURN $\mathbf{X}(x;w)\leftarrow\log(1+\mathbf{M})$
  \end{algorithmic}
\end{algorithm}

Unlike TraVerse~\cite{Traverse}, which serializes burst sequences into textual
prompts for a language-oriented backbone, SCSM directly represents traffic
through numerical packet-count and temporal features. Specifically, SCSM
adapts the Windowed Traffic Counting Matrix (WTCM) representation introduced in
CountMamba~\cite{CountMamba} to a direction-only setting. Given the trace
$x=((t_1,d_1),\ldots,(t_L,d_L))$, WTCM divides the observation period into
$N$ nonoverlapping windows of length $w$ and represents each window $j$ as:
\begin{equation}
  \mathbf{m}_j = [n_j^{+}, n_j^{-}, g_j, q_j]^{\mathsf{T}},
  \label{eq:wtcm-window-feature}
\end{equation}
where $n_j^{+}$ and $n_j^{-}$ are the numbers of outgoing and incoming packets,
respectively. For each nonempty window, $g_j$ records the gap between the
current window and the most recent nonempty window, while $q_j$ counts the
packet clusters separated by inter-packet gaps greater than a threshold
$\delta$.

Algorithm~\ref{alg:direction-wtcm} summarizes the construction. The window
features form a matrix $\mathbf{M}(x;w)=[\mathbf{m}_1,\ldots,\mathbf{m}_N]
\in\mathbb{R}_{\geq 0}^{4\times N}$. Following CountMamba, we apply an
element-wise log transformation and use
$\mathbf{X}(x;w)=\log(1+\mathbf{M}(x;w))$ as the encoder input.
Making $w$ an explicit input allows temporal Scaling to change the aggregation
granularity during matrix construction, as detailed in
Section~\ref{sec:view-construction}.

\subsection{Traffic-Grounded Multi-Level View Construction}
\label{sec:view-construction}

SCSM constructs pretraining views that combine event-level fidelity with
structural diversity. Given the unlabeled corpus $\mathcal{U}$, it samples a
group $\mathcal{C}=(x_{i_1},\ldots,x_{i_M})$ of $M$ single-tab traces and feeds
the same group to two independently randomized construction branches to produce
semantically consistent views with different observable traffic patterns.
Each branch reuses observed packet events while introducing variations at
three complementary levels. Trace-level Segmentation and Combination vary the
global organization of packet fragments to create alternative composite traces.
Matrix-level temporal Scaling changes their observation granularity without
modifying the underlying packet events. Feature-level Masking randomly removes
temporal regions, encouraging the encoder to integrate evidence across the
observation horizon.

\subsubsection{Trace-Level Segmentation and Combination}

Given the sampled group
$\mathcal{C}=(x_{i_1},\ldots,x_{i_M})$, each view-construction branch converts
the $M$ constituent traces into a single composite trace $\bar{x}^{(v)}$.
This trace-level stage supplies the main structural diversity by reorganizing
segments within individual traces and combining fragments across traces.
By preserving packet directions and relative timings within local fragments,
the construction retains the local dynamics of observed packet events.

Intra-trace reorganization divides each constituent trace $x_{i_m}$ into
$K$ packet-contiguous segments. For each view, SCSM independently samples the
partition boundaries and the order in which the segments are arranged.
The segments are temporally rebased and concatenated in that order, producing
a reorganized trace of duration $D_m^{(v)}$. This reorganized trace serves as
the source for fragment selection in the subsequent inter-trace combination.

To combine these reorganized traces, SCSM selects one fragment from each trace
and arranges the $M$ fragments with possible overlaps or gaps. For view
$v\in\{1,2\}$, SCSM samples an overlap ratio:
\begin{equation*}
  r^{(v)}\sim\operatorname{Unif}[-r_{\max},r_{\max}],
  \qquad 0\leq r_{\max}\leq 1.
\end{equation*}
A positive value causes adjacent fragments to overlap, zero places them
consecutively, and a negative value inserts a gap. Let
$T^{(v)}=Nw^{(v)}$ denote the view-specific observation horizon, where $N$ is
the fixed number of WTCM columns and $w^{(v)}$ is the sampled window length
used for matrix-level temporal Scaling in the next step.
SCSM sets the maximum duration of each selected fragment to:
\begin{equation}
  T_{\mathrm{frag}}^{(v)} =
  \frac{T^{(v)}}{M-r^{(v)}(M-1)}.
  \label{eq:fragment-duration}
\end{equation}
When $D_m^{(v)}>T_{\mathrm{frag}}^{(v)}$, SCSM samples
$a_m^{(v)}\sim\operatorname{Unif}[0,D_m^{(v)}-T_{\mathrm{frag}}^{(v)}]$
and retains the packets in
$[a_m^{(v)},a_m^{(v)}+T_{\mathrm{frag}}^{(v)}]$. A shorter trace is retained in
full. Each selected fragment is then rebased to start at time zero.

SCSM randomly orders the $M$ fragments and places the first at time zero.
Each subsequent fragment is positioned relative to the end of the previous
one: it starts earlier to create an overlap when $r^{(v)}>0$, exactly at the
previous fragment's end when $r^{(v)}=0$, or later to insert a gap when
$r^{(v)}<0$. The overlap or gap duration is $|r^{(v)}|$ times the shorter
actual duration of the two adjacent fragments. After positioning all
fragments, SCSM merges their packets and stably sorts them by timestamp to
obtain $\bar{x}^{(v)}$.
Because $\bar{x}^{(1)}$ and $\bar{x}^{(2)}$ are both derived from every member
of the same group $\mathcal{C}$, they share the same latent website composition
while exhibiting different global temporal arrangements. This correspondence
is established from the shared source group without using website annotations.

\subsubsection{Matrix-Level Temporal Scaling}

Traffic collected under different network paths or client configurations may
exhibit temporal stretching or compression~\cite{Swallow}. SCSM introduces this variation by
changing how a composite trace is temporally aggregated, without modifying its
underlying packet events. For each view, SCSM samples the window length
$w^{(v)}$, which also determines the observation horizon $T^{(v)}$ used above:
\begin{equation}
  w^{(v)} \sim \operatorname{Unif}[w_{\min},w_{\max}],
  \label{eq:temporal-scaling}
\end{equation}
It then constructs
$\mathbf{X}^{(v)}=\mathbf{X}(\bar{x}^{(v)};w^{(v)})$ using
Algorithm~\ref{alg:direction-wtcm}. Increasing $w^{(v)}$ aggregates a longer
interval into each column and compresses the traffic pattern, whereas
decreasing it spreads the pattern across more columns. The output remains a
$4\times N$ direction-only WTCM at every scale, enabling all views to share the
same encoder.

\subsubsection{Feature-Level Masking}

Finally, SCSM randomly masks complete WTCM columns, thereby removing temporal
regions from the matrix view and encouraging the encoder to integrate evidence
across the observation horizon. Specifically, it uniformly samples the
following column set without replacement:
\begin{equation*}
  \mathcal{J}^{(v)}\subseteq\{1,\ldots,N\},
  \qquad |\mathcal{J}^{(v)}|=\lfloor\rho N\rfloor,
\end{equation*}
where $\rho$ is the masking ratio, and defines $b_j^{(v)}=0$ for
$j\in\mathcal{J}^{(v)}$ and $b_j^{(v)}=1$ otherwise. The final view is then
computed as:
\begin{equation}
  \widetilde{\mathbf{X}}^{(v)}
  =\mathbf{X}^{(v)}\odot
  \left(\mathbf{1}_4(\mathbf{b}^{(v)})^{\mathsf{T}}\right),
  \label{eq:feature-masking}
\end{equation}
where $\odot$ denotes element-wise multiplication, $\mathbf{1}_4$ is a
four-dimensional all-ones vector, and
$\mathbf{b}^{(v)}=[b_1^{(v)},\ldots,b_N^{(v)}]^{\mathsf{T}}$ is the binary
temporal mask.

Overall, each branch implements the following transformation:
\begin{equation*}
  \widetilde{\mathbf{X}}^{(v)}
  =\mathcal{A}_{\boldsymbol{\xi}^{(v)}}(\mathcal{C}),
\end{equation*}
where $\boldsymbol{\xi}^{(v)}$ collects its segmentation, crop, combination,
scale, and masking variables. Reusing observed fragments provides event-level
grounding, whereas independently sampling these variables produces diversity
in global organization, temporal granularity, and visible regions. Because the
two branches share the same group $\mathcal{C}$, their final views retain the
same latent website composition despite these structural differences. The pair
$(\widetilde{\mathbf{X}}^{(1)},\widetilde{\mathbf{X}}^{(2)})$ can therefore be
used for contrastive pretraining without website annotations.

\subsection{State-Space Encoder and Contrastive Pretraining}
\label{sec:encoder-pretraining}

SCSM maps the paired views constructed in
Section~\ref{sec:view-construction} into traffic embeddings using a shared
encoder $f_{\boldsymbol{\theta}}$, which is trained with a cross-view
contrastive objective without website annotations.

\subsubsection{State-Space Traffic Encoding}

The encoder follows the state-space-oriented architecture of
CountMamba~\cite{CountMamba}, adapted to the four-channel WTCM input.
A $4\times 1$ convolution first projects each window of
$\widetilde{\mathbf{X}}^{(v)}\in\mathbb{R}^{4\times N}$ into a
$d$-dimensional embedding. Four causal convolutional blocks with temporal
max pooling then extract local traffic patterns and reduce the sequence
length by a factor of six. After adding sinusoidal positional embeddings,
a stack of Mamba2 blocks~\cite{Mamba2} captures temporal dependencies.

\subsubsection{View-Level Representation}

The encoder's temporal outputs are averaged across temporal positions
to obtain a single vector representation
$\mathbf{z}_i^{(v)}=f_{\boldsymbol{\theta}}(\widetilde{\mathbf{X}}_i^{(v)})$
for each view. This representation is used directly in the contrastive
objective, without a separate projection head.

\subsubsection{Cross-View Contrastive Objective}

For a batch of $B$ source groups $\{\mathcal{C}_i\}_{i=1}^{B}$, the shared
encoder produces paired embeddings
$\{(\mathbf{z}_i^{(1)},\mathbf{z}_i^{(2)})\}_{i=1}^{B}$.
For each anchor in one view, the corresponding embedding in the other view
serves as the positive, while the remaining embeddings in that view serve
as negatives.

For $\mathbf{z}_i^{(1)}$ as the anchor, the cross-view loss is:
\begin{equation}
  \ell_i^{1\rightarrow 2}
  = -\log
  \frac{\exp\!\left(\operatorname{sim}(\mathbf{z}_i^{(1)},
  \mathbf{z}_i^{(2)})/\gamma\right)}
  {\sum_{j=1}^{B}\exp\!\left(\operatorname{sim}(\mathbf{z}_i^{(1)},
  \mathbf{z}_j^{(2)})/\gamma\right)},
  \label{eq:cross-view-loss}
\end{equation}
where $\operatorname{sim}(\cdot,\cdot)$ denotes cosine similarity and
$\gamma>0$ is the temperature parameter. The reverse loss
$\ell_i^{2\rightarrow 1}$ is obtained by exchanging the two branches, and
the pretraining objective averages both directions over the batch:
\begin{equation}
  \mathcal{L}_{\mathrm{pre}}
  = \frac{1}{2B}\sum_{i=1}^{B}
    \left(\ell_i^{1\rightarrow 2}+\ell_i^{2\rightarrow 1}\right).
  \label{eq:symmetric-pretraining}
\end{equation}
Minimizing $\mathcal{L}_{\mathrm{pre}}$ updates the shared encoder to align
views of the same source group despite changes in global fragment
organization, temporal granularity, and visible regions. This connects
the traffic-grounded view construction to representation learning,
yielding a pretrained encoder for downstream few-shot adaptation.

\subsection{Few-Shot Adaptation}
\label{sec:scsm-adaptation}

For a target task $\tau$, SCSM appends a single linear classification layer
$h_{\boldsymbol{\phi}_{\tau}}$ to the pretrained encoder and adapts the
resulting model on the labeled support set $\mathcal{S}_{\tau}$, as described
in Section~\ref{sec:threat-learning-paradigm}. Each training view is
constructed with a randomly sampled window length
$w\sim\operatorname{Unif}[w_{\min},w_{\max}]$.
This applies the matrix-level temporal Scaling mechanism from
Section~\ref{sec:view-construction} to adaptation, exposing the classifier
to target traces at varying temporal resolutions while preserving their
website labels.

At inference, SCSM aggregates evidence across temporal resolutions through
a multi-scale ensemble. It selects $R\geq 2$ equally spaced window lengths
over the same interval:
\begin{equation*}
  w_r=w_{\min}+\frac{r-1}{R-1}(w_{\max}-w_{\min}),
  \quad r=1,\ldots,R.
\end{equation*}
For a query trace $x$, Algorithm~\ref{alg:direction-wtcm} constructs the
corresponding views $\{\mathbf{X}(x;w_r)\}_{r=1}^{R}$. Each view represents
the same packet events at a different aggregation granularity and retains
the $4\times N$ input shape.

The adapted encoder and linear head share their parameters across all
views and produce the class logits:
\begin{equation*}
  \mathbf{a}_{\tau}(x;w_r)=h_{\boldsymbol{\phi}_{\tau}}
  (f_{\boldsymbol{\theta}}(\mathbf{X}(x;w_r))).
\end{equation*}
The final prediction is obtained by averaging the logits across views
and selecting the website with the highest score:
\begin{equation}
  \widehat{y}_{\tau}(x)
  =\arg\max_{c\in\mathcal{Y}_{\tau}}
  \left[\frac{1}{R}\sum_{r=1}^{R}
  \mathbf{a}_{\tau}(x;w_r)\right]_c.
  \label{eq:multiscale-prediction}
\end{equation}
where $\mathcal{Y}_{\tau}$ is the target website set and $[\cdot]_c$
denotes the score for class $c$. This ensemble integrates classification
evidence across the temporal scale range used during adaptation.

\section{Evaluation}
\label{sec:evaluation}
We first pretrain SCSM on unlabeled Tor traffic through a self-supervised
objective to learn a traffic-native foundation model for WF. Using the
resulting pretrained checkpoints, we evaluate how the learned
representations support adaptation across
scenarios, domains, and tasks with limited labels. Finally, we examine
component contributions and parameter sensitivity.

\subsection{Experimental Setup}
\label{sec:experimental-setup}

\subsubsection{Datasets}
We pretrain separate SCSM foundation models on unlabeled traffic from
GTT23~\cite{GTT23} and Swallow D1~\cite{Swallow}. Each pretrained
checkpoint is then reused for cross-scenario adaptation within its
source dataset, as follows.
\begin{itemize}
\item \emph{GTT23.} This dataset contains real-user Tor traffic spanning
13 weeks. We use all traffic from the first 7 weeks for unlabeled
pretraining. To evaluate adaptation to temporal drift, we select 200
eligible websites and independently adapt the pretrained encoder in
each of weeks 8--13 using 10 labeled traces per website, with five
disjoint traces per website for testing.

\item \emph{Swallow.} These datasets comprise seven scenarios, D1--D7,
each with 100 websites and 100 traces per website. The pretraining
scenario, D1, uses Tor Browser and a Chicago guard relay in July 2024.
D2--D5 change the guard location to Singapore, London,
Johannesburg, and Mumbai, respectively; D6 changes the collection time
to October 2024; D7 changes the browser to Chrome. We transfer the
D1-pretrained encoder to D2--D7 using ten labeled traces per website
(10-shot), with the remaining 90 traces used for testing.
\end{itemize}

To assess reuse beyond the pretraining datasets, we evaluate
cross-domain transfer in four target settings. The models pretrained on
GTT23 and Swallow D1 are each fine-tuned independently on every target
dataset, starting from their respective pretrained checkpoints.
Single-tab adaptation uses ten
labeled traces per website (10-shot), while each multi-tab setting
uses 3,000 labeled traces in total. The remaining traces are used for
testing.

\begin{itemize}
\item \emph{Closed-world.} We evaluate website classification on
DFSet~\cite{DF} (95 websites, 1,000 traces each), simulated
W-TSet~\cite{walkietalkie} (100 websites, 900 traces each),
k-NNSet~\cite{wang2014} (100 websites, 90 traces each), and
R-PSet~\cite{mei2025hpetc} (265 websites, 100 traces each).

\item \emph{Open-world.} Open-World DFSet combines DFSet with 40,000
unmonitored traces, each from a distinct website. We fine-tune on ten
labeled traces per monitored website and an equal number of unmonitored
traces, using the rest for testing.

\item \emph{Multi-tab.} We assess transfer from single-tab to multi-tab traffic using
ARESSet~\cite{ARES}, covering 2--5 tabs and 100 monitored websites.
Both closed-world and open-world versions of each tab setting are evaluated.

\item \emph{Defended.} We use Defended DFSet to evaluate transfer to traffic modified by WF
defenses, with adaptation performed separately under each defense.
\end{itemize}

\subsubsection{Implementation Details}
We conduct all experiments on four NVIDIA RTX 4090 GPUs using
PyTorch 2.5.1 and CUDA 12.1. SCSM uses $4\times N$ WTCMs with
$N=2700$ and $7200$ for single-tab tasks and multi-tab transfer,
respectively. The encoder has embedding dimension $d=512$ and three
Mamba2 blocks with a maximum drop-path rate of 0.2; its local causal
convolutions use kernel size five and dropout 0.1 after temporal
pooling. For view construction during pretraining, we set the segment
count $K=2$, sample the source-trace count $M$ uniformly from
$\{2,3,4,5\}$, and set the maximum overlap ratio $r_{\max}=0.5$.
Window lengths are sampled from $[w_{\min},w_{\max}]=[5,50]$ ms,
with masking ratio $\rho=0.5$ and cluster threshold $\delta=0.1w$.
The contrastive temperature is $\gamma=0.07$.

We use AdamW with weight decay 0.05 for SCSM pretraining and fine-tuning.
Pretraining runs for five epochs on GTT23 or 1,500 epochs on Swallow D1,
with learning rate $2\times10^{-3}$, 1,000 warmup steps, and cosine
decay to zero. Per-GPU batch sizes are 128 and 32 for SCSM and
SCSM-multi, respectively. The pretrained encoder and a linear
classifier are then fine-tuned for 200 epochs with batch size 128,
learning rate $10^{-4}$, label smoothing 0.1, and gradient clipping
at a maximum norm of 1. Fine-tuning and inference retain the pretraining
window-length range. Results are reported using the final fine-tuned
checkpoint with $R=5$ inference scales.

We reproduce the baselines following the hyperparameters reported in
their original papers. For fair comparison, transferable baselines use
the same pretraining data and epoch budgets as SCSM, and all methods
share the same fine-tuning and test splits, label budgets, and
fine-tuning epochs. For the window-based baselines RF and CountMamba,
we match SCSM's window count, using $N=2700$ for single-tab tasks and
$N=7200$ for multi-tab tasks.

\subsubsection{WF Attacks and Defenses}
We compare with three transferable attacks: NetCLR~\cite{NetCLR},
Swallow~\cite{Swallow}, and TraVerse~\cite{Traverse}. The comparison
also includes eight supervised attacks:
AWF~\cite{AWF}, DF~\cite{DF}, Tik-Tok~\cite{Tik-Tok},
Var-CNN~\cite{Var-CNN}, RF~\cite{RF}, CountMamba~\cite{CountMamba},
TMWF~\cite{TMWF}, and ARES~\cite{ARES}. To evaluate transfer to defended
traffic, we consider six defenses: WTF-PAD~\cite{wtfpad}, FRONT~\cite{front},
RegulaTor~\cite{regulator}, Tamaraw~\cite{tamaraw},
Palette~\cite{Palette}, and TrafficSliver~\cite{trafficsliver}.

\subsubsection{Evaluation Metrics}
We report accuracy and macro-F1 for single-tab classification; for GTT23,
we follow TraVerse~\cite{Traverse} and report Accuracy@3.

For Open-World DFSet, we report $r$-precision and recall with $r=20$
following Holmes~\cite{Holmes}, where $r$ denotes the ratio of
unmonitored to monitored visits. Specifically,
\begin{equation}
  \pi_r=\frac{\mathrm{TPR}}
  {\mathrm{TPR}+\mathrm{WPR}+r\,\mathrm{FPR}},
  \label{eq:r-precision}
\end{equation}
where TPR (recall) and WPR are the fractions of monitored traces assigned the
correct and an incorrect monitored website, respectively, and FPR is
the fraction of unmonitored traces accepted as monitored. We sweep
confidence thresholds over $[0,1]$ to obtain
$20$-precision--recall curves.

For multi-tab tasks, we report P@$k$ and MAP@$k$ following
ARES~\cite{ARES}, with $k$ equal to the number of tabs. Results are
reported separately for each target
week, scenario, and closed-world or open-world setting.

\subsection{Temporal Drift Evaluation}
\label{sec:temporal-drift-evaluation}
\begin{figure*}[!t]
  \centering
  \input{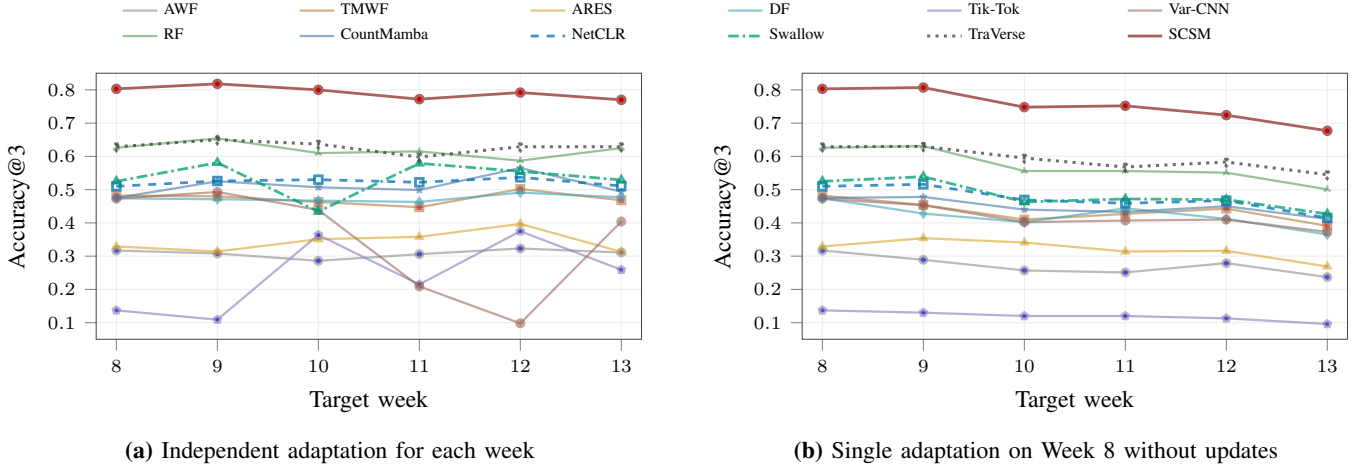}
  \pgfplotslegendfromname{gttTemporalLegend}
  \vspace{-5pt}

  \begin{minipage}[t]{0.485\textwidth}
    \vspace{0pt}
    \centering
\begin{tikzpicture}
\begin{axis}[
  width=\linewidth,
  height=0.58\linewidth,
  xmin=7.8, xmax=13.2,
  ymin=0.05, ymax=0.85,
  xtick={8,9,10,11,12,13},
  ytick={0.1,0.2,0.3,0.4,0.5,0.6,0.7,0.8},
  xlabel={Target week},
  ylabel={Accuracy@3},
  tick pos=left,
  tick align=outside,
  tick label style={font=\scriptsize},
  label style={font=\small},
  grid=major,
  grid style={draw=black!7, line width=0.20pt},
  axis line style={black!70, line width=0.4pt},
  legend image post style={xscale=0.76, mark=none},
  legend to name=gttTemporalLegend,
  legend columns=6,
  legend style={
    draw=none, fill=none,
    font=\fontsize{6.0}{6.4}\selectfont,
    legend cell align=left,
    /tikz/every even column/.append style={column sep=36pt},
    /tikz/every row/.append style={row sep=-0.4pt}
  },
  clip mode=individual
]
\addplot+[
  color=plotAWF, opacity=0.60, solid, line width=0.85pt,
  mark=*, mark size=1.45pt
] coordinates {(8,0.317000) (9,0.308000) (10,0.286000) (11,0.306000) (12,0.323000) (13,0.311000)};
\addlegendentry{AWF}
\addplot+[
  color=plotTMWF, opacity=0.60, solid, line width=0.85pt,
  mark=square*, mark size=1.45pt
] coordinates {(8,0.482000) (9,0.480000) (10,0.463000) (11,0.447000) (12,0.502000) (13,0.468000)};
\addlegendentry{TMWF}
\addplot+[
  color=plotARES, opacity=0.60, solid, line width=0.85pt,
  mark=triangle*, mark size=1.45pt
] coordinates {(8,0.329000) (9,0.314000) (10,0.351000) (11,0.358000) (12,0.397000) (13,0.313000)};
\addlegendentry{ARES}
\addplot+[
  color=plotDF, opacity=0.60, solid, line width=0.85pt,
  mark=diamond*, mark size=1.45pt
] coordinates {(8,0.474000) (9,0.471000) (10,0.467000) (11,0.463000) (12,0.491000) (13,0.477000)};
\addlegendentry{DF}
\addplot+[
  color=plotTikTok, opacity=0.60, solid, line width=0.85pt,
  mark=pentagon*, mark size=1.45pt
] coordinates {(8,0.137000) (9,0.109000) (10,0.364000) (11,0.215000) (12,0.375000) (13,0.259000)};
\addlegendentry{Tik-Tok}
\addplot+[
  color=plotVarCNN, opacity=0.60, solid, line width=0.85pt,
  mark=otimes*, mark size=1.45pt
] coordinates {(8,0.473000) (9,0.493000) (10,0.439000) (11,0.209000) (12,0.098000) (13,0.404000)};
\addlegendentry{Var-CNN}
\addplot+[
  color=plotRF, opacity=0.60, solid, line width=0.85pt,
  mark=Mercedes star, mark size=1.45pt
] coordinates {(8,0.626000) (9,0.654000) (10,0.610000) (11,0.615000) (12,0.587000) (13,0.625000)};
\addlegendentry{RF}
\addplot+[
  color=plotCountMamba, opacity=0.60, solid, line width=0.85pt,
  mark=star, mark size=1.45pt
] coordinates {(8,0.475000) (9,0.525000) (10,0.507000) (11,0.499000) (12,0.565000) (13,0.494000)};
\addlegendentry{CountMamba}
\addplot+[
  color=plotNetCLR, opacity=0.76, dashed, line width=1.05pt,
  mark=square, mark size=1.45pt
] coordinates {(8,0.510000) (9,0.526000) (10,0.530000) (11,0.522000) (12,0.537000) (13,0.511000)};
\addlegendentry{NetCLR}
\addplot+[
  color=plotSwallow, opacity=0.76, dash pattern=on 4pt off 1.4pt on 1pt off 1.4pt, line width=1.05pt,
  mark=triangle, mark size=1.45pt
] coordinates {(8,0.525000) (9,0.581000) (10,0.434000) (11,0.579000) (12,0.555000) (13,0.529000)};
\addlegendentry{Swallow}
\addplot+[
  color=plotTraVerse, opacity=0.76, dotted, line width=1.15pt,
  mark=diamond, mark size=1.45pt
] coordinates {(8,0.629000) (9,0.650000) (10,0.637000) (11,0.598000) (12,0.629000) (13,0.629000)};
\addlegendentry{TraVerse}
\addplot+[
  color=plotSCSM, opacity=1.0, solid, line width=1.05pt,
  mark=*, mark size=1.45pt
] coordinates {(8,0.803000) (9,0.818000) (10,0.800000) (11,0.772000) (12,0.792000) (13,0.770000)};
\addlegendentry{SCSM}
\end{axis}
\end{tikzpicture}
    \vspace{4pt}

    \small\textbf{(a)} Independent adaptation for each week
  \end{minipage}
  \hfill
  \begin{minipage}[t]{0.485\textwidth}
    \vspace{0pt}
    \centering
\begin{tikzpicture}
\begin{axis}[
  width=\linewidth,
  height=0.58\linewidth,
  xmin=7.8, xmax=13.2,
  ymin=0.05, ymax=0.85,
  xtick={8,9,10,11,12,13},
  ytick={0.1,0.2,0.3,0.4,0.5,0.6,0.7,0.8},
  xlabel={Target week},
  ylabel={Accuracy@3},
  tick pos=left,
  tick align=outside,
  tick label style={font=\scriptsize},
  label style={font=\small},
  grid=major,
  grid style={draw=black!7, line width=0.20pt},
  axis line style={black!70, line width=0.4pt},
  legend image post style={xscale=0.76, mark=none},
  clip mode=individual
]
\addplot+[
  color=plotAWF, opacity=0.60, solid, line width=0.85pt,
  mark=*, mark size=1.45pt
] coordinates {(8,0.317000) (9,0.289000) (10,0.257000) (11,0.251000) (12,0.279000) (13,0.237000)};
\addplot+[
  color=plotTMWF, opacity=0.60, solid, line width=0.85pt,
  mark=square*, mark size=1.45pt
] coordinates {(8,0.482000) (9,0.454000) (10,0.410000) (11,0.427000) (12,0.443000) (13,0.391000)};
\addplot+[
  color=plotARES, opacity=0.60, solid, line width=0.85pt,
  mark=triangle*, mark size=1.45pt
] coordinates {(8,0.329000) (9,0.354000) (10,0.341000) (11,0.314000) (12,0.316000) (13,0.269000)};
\addplot+[
  color=plotDF, opacity=0.60, solid, line width=0.85pt,
  mark=diamond*, mark size=1.45pt
] coordinates {(8,0.474000) (9,0.428000) (10,0.402000) (11,0.444000) (12,0.412000) (13,0.365000)};
\addplot+[
  color=plotTikTok, opacity=0.60, solid, line width=0.85pt,
  mark=pentagon*, mark size=1.45pt
] coordinates {(8,0.137000) (9,0.130000) (10,0.120000) (11,0.120000) (12,0.113000) (13,0.096000)};
\addplot+[
  color=plotVarCNN, opacity=0.60, solid, line width=0.85pt,
  mark=otimes*, mark size=1.45pt
] coordinates {(8,0.473000) (9,0.453000) (10,0.402000) (11,0.407000) (12,0.410000) (13,0.373000)};
\addplot+[
  color=plotRF, opacity=0.60, solid, line width=0.85pt,
  mark=Mercedes star, mark size=1.45pt
] coordinates {(8,0.626000) (9,0.631000) (10,0.556000) (11,0.556000) (12,0.551000) (13,0.501000)};
\addplot+[
  color=plotCountMamba, opacity=0.60, solid, line width=0.85pt,
  mark=star, mark size=1.45pt
] coordinates {(8,0.475000) (9,0.478000) (10,0.440000) (11,0.433000) (12,0.450000) (13,0.412000)};
\addplot+[
  color=plotNetCLR, opacity=0.76, dashed, line width=1.05pt,
  mark=square, mark size=1.45pt
] coordinates {(8,0.510000) (9,0.516000) (10,0.469000) (11,0.459000) (12,0.467000) (13,0.415000)};
\addplot+[
  color=plotSwallow, opacity=0.76, dash pattern=on 4pt off 1.4pt on 1pt off 1.4pt, line width=1.05pt,
  mark=triangle, mark size=1.45pt
] coordinates {(8,0.525000) (9,0.539000) (10,0.463000) (11,0.472000) (12,0.470000) (13,0.427000)};
\addplot+[
  color=plotTraVerse, opacity=0.76, dotted, line width=1.15pt,
  mark=diamond, mark size=1.45pt
] coordinates {(8,0.629000) (9,0.630000) (10,0.595000) (11,0.568000) (12,0.583000) (13,0.545000)};
\addplot+[
  color=plotSCSM, opacity=1.0, solid, line width=1.05pt,
  mark=*, mark size=1.45pt
] coordinates {(8,0.803000) (9,0.807000) (10,0.748000) (11,0.752000) (12,0.724000) (13,0.677000)};
\end{axis}
\end{tikzpicture}
    \vspace{4pt}

    \small\textbf{(b)} Single adaptation on Week 8 without updates
  \end{minipage}
  \caption{Temporal-drift performance on GTT23 in the 10-shot setting.
  (a) Accuracy@3 when each target week is independently fine-tuned from the
  same pretrained checkpoint. (b) Accuracy@3 after adaptation on Week~8,
  with the resulting model evaluated on all subsequent weeks without updates.}
  \label{fig:gtt23-temporal-drift}
\end{figure*}

\label{sec:cross-scenario-evaluation}
\begin{table*}[!t]
  \caption{Accuracy and Macro-F1 of WF Attacks under 10-Shot
  Cross-Scenario Transfer.}
  \label{tab:swallow-cross-scenario}
  \centering
  \footnotesize
  \setlength{\tabcolsep}{4.50pt}
  \renewcommand{\arraystretch}{1.08}
  \begin{tabular}{@{}l*{7}{cc}@{}}
    \toprule
    \multirow{3}{*}{Method} & \multicolumn{10}{c}{Guard-relay location}
    & \multicolumn{2}{c}{Concept drift}
    & \multicolumn{2}{c}{Browser} \\
    \cmidrule(lr){2-11}
    \cmidrule(lr){12-13}\cmidrule(l){14-15}
    & \multicolumn{2}{c}{D2} & \multicolumn{2}{c}{D3}
    & \multicolumn{2}{c}{D4} & \multicolumn{2}{c}{D5}
    & \multicolumn{2}{c}{Average}
    & \multicolumn{2}{c}{D6} & \multicolumn{2}{c}{D7} \\
    \cmidrule(lr){2-3}\cmidrule(lr){4-5}\cmidrule(lr){6-7}
    \cmidrule(lr){8-9}\cmidrule(lr){10-11}\cmidrule(lr){12-13}
    \cmidrule(l){14-15}
    & Acc. & F1 & Acc. & F1 & Acc. & F1 & Acc. & F1
      & Acc. & F1 & Acc. & F1 & Acc. & F1 \\
    \midrule
    AWF        & 0.822 & 0.819 & 0.826 & 0.828 & 0.830 & 0.828 & 0.795 & 0.793 & 0.818 & 0.817 & 0.840 & 0.839 & 0.251 & 0.245 \\
    TMWF       & 0.890 & 0.889 & 0.870 & 0.869 & 0.886 & 0.885 & 0.869 & 0.867 & 0.879 & 0.877 & 0.904 & 0.902 & 0.643 & 0.640 \\
    ARES       & 0.536 & 0.521 & 0.465 & 0.448 & 0.553 & 0.545 & 0.513 & 0.498 & 0.517 & 0.503 & 0.527 & 0.514 & 0.127 & 0.113 \\
    DF         & 0.909 & 0.908 & 0.892 & 0.891 & 0.915 & 0.915 & 0.895 & 0.894 & 0.903 & 0.902 & 0.926 & 0.925 & 0.624 & 0.616 \\
    Tik-Tok    & 0.903 & 0.903 & 0.894 & 0.893 & 0.918 & 0.918 & 0.904 & 0.903 & 0.905 & 0.904 & 0.923 & 0.922 & 0.613 & 0.610 \\
    Var-CNN    & 0.765 & 0.761 & 0.773 & 0.770 & 0.847 & 0.846 & 0.676 & 0.665 & 0.765 & 0.760 & 0.708 & 0.701 & 0.336 & 0.315 \\
    RF         & 0.894 & 0.894 & 0.878 & 0.881 & 0.919 & 0.919 & 0.902 & 0.902 & 0.898 & 0.899 & 0.884 & 0.883 & 0.455 & 0.456 \\
    CountMamba & 0.879 & 0.878 & 0.825 & 0.824 & 0.873 & 0.873 & 0.881 & 0.880 & 0.864 & 0.864 & 0.869 & 0.869 & 0.411 & 0.412 \\
    \midrule
    NetCLR   & 0.921 & 0.921 & 0.921 & 0.921 & 0.931 & 0.931 & 0.921 & 0.922 & 0.924 & 0.924 & 0.932 & 0.931 & 0.549 & 0.547 \\
    Swallow  & 0.896 & 0.896 & 0.868 & 0.873 & 0.909 & 0.909 & 0.899 & 0.900 & 0.893 & 0.895 & 0.883 & 0.884 & 0.407 & 0.406 \\
    TraVerse & 0.900 & 0.899 & 0.913 & 0.913 & 0.902 & 0.902 & 0.890 & 0.888 & 0.902 & 0.901 & 0.876 & 0.871 & 0.278 & 0.253 \\
    \midrule
    \textbf{SCSM} & \textbf{0.961} & \textbf{0.961}
      & \textbf{0.961} & \textbf{0.962}
      & \textbf{0.971} & \textbf{0.972}
      & \textbf{0.969} & \textbf{0.970}
      & \textbf{0.966} & \textbf{0.966}
      & \textbf{0.970} & \textbf{0.970}
      & \textbf{0.699} & \textbf{0.697} \\
    \bottomrule
  \end{tabular}
\end{table*}

We evaluate temporal drift on the same 200 websites over Weeks~8--13,
with transferable attacks pretrained on unlabeled traffic from Weeks~1--7.
Under \emph{independent adaptation}, current-week labels are available:
each attack is trained or fine-tuned on that week's 10-shot support set,
with transferable attacks initialized from their original pretrained
checkpoints. Under \emph{single adaptation}, no new labels are available
after Week~8: each attack uses only the Week~8 10-shot support set and is
evaluated on later weeks without updates.

\textbf{Independent adaptation.}
Figure~\ref{fig:gtt23-temporal-drift}(a) shows that SCSM achieves the highest
Accuracy@3 in every target week, ranging from 77.0\% to 81.8\%. Its six-week
average of 79.3\% exceeds the strongest supervised baseline, RF, and the
strongest transferable baseline, TraVerse, by 17.3\% and 16.4\%, respectively.
Pretrained methods (i.e., NetCLR, Swallow, TraVerse, and SCSM) gain an
advantage over supervised methods by reusing historical traffic representations
for few-shot adaptation as website content evolves.
SCSM further enhances transfer through Segmentation and
Combination, which reorganize and combine observed traffic fragments to
simulate more diverse website traffic patterns. Learning from these diverse
patterns helps the encoder adapt to traffic changes caused by website updates,
contributing to SCSM's superior performance under temporal drift.

\textbf{Single adaptation without updates.}
Without updates, model accuracy generally declines as temporal drift
accumulates, yet SCSM remains the best-performing method in every target week
(Fig.~\ref{fig:gtt23-temporal-drift}(b)). Its Accuracy@3 falls from 80.3\%
in Week~8 to 67.7\% in Week~13, still well above TraVerse's 54.5\%.
Diverse pretraining views help SCSM learn features that generalize beyond
the traffic patterns observed during adaptation, sustaining its advantage
without new labels.

\begin{table*}[!t]
  \caption{Accuracy and Macro-F1 of WF Attacks under 10-Shot
  Closed-World Cross-Domain Transfer.}
  \label{tab:closed-world-transfer}
  \centering
  \footnotesize
  \setlength{\tabcolsep}{6.5pt}
  \renewcommand{\arraystretch}{1.15}
  \begin{tabular}{@{}ll*{5}{cc}@{}}
    \toprule
    \multirow{2}{*}{Method} & \multirow{2}{*}{Source}
      & \multicolumn{2}{c}{DFSet} & \multicolumn{2}{c}{W-TSet}
      & \multicolumn{2}{c}{k-NNSet} & \multicolumn{2}{c}{R-PSet}
      & \multicolumn{2}{c}{Average} \\
    \cmidrule(lr){3-4}\cmidrule(lr){5-6}
    \cmidrule(lr){7-8}\cmidrule(lr){9-10}\cmidrule(l){11-12}
    & & Acc. & F1 & Acc. & F1 & Acc. & F1 & Acc. & F1
      & Acc. & F1 \\
    \midrule
    AWF        & Target only & 0.684 & 0.682 & 0.193 & 0.189 & 0.251 & 0.362 & 0.360 & 0.341 & 0.372 & 0.394 \\
    TMWF       & Target only & 0.761 & 0.755 & 0.333 & 0.324 & 0.279 & 0.351 & 0.385 & 0.368 & 0.440 & 0.450 \\
    ARES       & Target only & 0.458 & 0.444 & 0.251 & 0.212 & 0.151 & 0.142 & 0.374 & 0.359 & 0.309 & 0.289 \\
    DF         & Target only & 0.824 & 0.820 & 0.377 & 0.369 & 0.305 & 0.429 & 0.435 & 0.415 & 0.485 & 0.508 \\
    Tik-Tok    & Target only & 0.806 & 0.802 & 0.596 & 0.590 & 0.300 & 0.421 & 0.440 & 0.420 & 0.536 & 0.558 \\
    Var-CNN    & Target only & 0.688 & 0.685 & 0.429 & 0.415 & 0.056 & 0.012 & 0.307 & 0.293 & 0.370 & 0.351 \\
    RF         & Target only & 0.835 & 0.840 & 0.624 & 0.618 & 0.732 & 0.693 & 0.540 & 0.548 & 0.683 & 0.675 \\
    CountMamba & Target only & 0.791 & 0.789 & 0.720 & 0.716 & 0.416 & 0.508 & 0.533 & 0.522 & 0.615 & 0.634 \\
    \midrule
    NetCLR   & GTT23      & 0.752 & 0.751 & 0.380 & 0.372 & 0.356 & 0.451 & 0.459 & 0.448 & 0.487 & 0.506 \\
    NetCLR   & Swallow D1 & 0.817 & 0.813 & 0.368 & 0.361 & 0.417 & 0.519 & 0.458 & 0.450 & 0.515 & 0.536 \\
    Swallow  & GTT23      & 0.642 & 0.642 & 0.438 & 0.441 & 0.466 & 0.441 & 0.251 & 0.254 & 0.449 & 0.445 \\
    Swallow  & Swallow D1 & 0.765 & 0.766 & 0.466 & 0.471 & 0.585 & 0.574 & 0.281 & 0.285 & 0.524 & 0.524 \\
    TraVerse & GTT23      & 0.265 & 0.235 & 0.296 & 0.272 & 0.290 & 0.259 & 0.122 & 0.088 & 0.243 & 0.214 \\
    TraVerse & Swallow D1 & 0.348 & 0.315 & 0.370 & 0.351 & 0.272 & 0.249 & 0.104 & 0.066 & 0.274 & 0.245 \\
    \midrule
    \textbf{SCSM} & GTT23
      & \underline{0.864} & \underline{0.863}
      & \underline{0.769} & \underline{0.769}
      & \textbf{0.862} & \textbf{0.812}
      & \underline{0.711} & \underline{0.706}
      & \underline{0.802} & \underline{0.788} \\
    \textbf{SCSM} & Swallow D1
      & \textbf{0.908} & \textbf{0.908}
      & \textbf{0.823} & \textbf{0.824}
      & \underline{0.849} & \underline{0.803}
      & \textbf{0.721} & \textbf{0.717}
      & \textbf{0.825} & \textbf{0.813} \\
    \bottomrule
  \end{tabular}
\end{table*}

\subsection{Cross-Scenario Transfer Evaluation}
We evaluate cross-scenario transfer on Swallow while keeping the 100
monitored websites fixed. The target scenarios vary guard-relay location
(D2--D5), collection time (D6), or browser (D7). Each transferable attack
is pretrained on unlabeled D1 traffic and independently adapted to each
target scenario using ten labeled traces per website.

Changes in guard-relay location (D2--D5) alter network latency, stretching
or compressing traffic patterns. Existing pretrained baselines offer only
limited gains over supervised methods (Table~\ref{tab:swallow-cross-scenario});
for example, NetCLR averages 92.4\% accuracy, compared with 90.5\% for
Tik-Tok. In contrast, SCSM uses Temporal Scaling to vary WTCM observation
granularity, emulating traffic patterns under different network delays
while reducing dependence on a fixed timing pattern. SCSM thus achieves
the highest accuracy across all four guard locations, averaging 96.6\%.

For temporal drift in D6, SCSM achieves the best performance,
consistent with the results on GTT23.

In D7, the browser changes from Tor Browser to Chrome. Existing pretrained
baselines transfer less effectively under this shift: NetCLR, Swallow, and
TraVerse achieve accuracies of 54.9\%, 40.7\%, and 27.8\%, respectively,
all below TMWF's 64.3\%. In contrast, SCSM achieves the highest accuracy
of 69.9\%, demonstrating that its pretrained model remains effective
for few-shot adaptation across browsers.

\subsection{Cross-Domain Transfer Evaluation}
\label{sec:cross-domain-evaluation}

To assess transferability beyond the pretraining datasets, we independently
adapt models pretrained on GTT23 or Swallow D1 to target datasets with
limited labeled supervision, covering closed-world classification,
open-world recognition, defended traffic, and multi-tab browsing.

\subsubsection{Closed-World Evaluation}
We evaluate closed-world transfer on DFSet, W-TSet, k-NNSet, and
R-PSet. For each target dataset, transferable attacks are independently
adapted from their GTT23- or Swallow D1-pretrained checkpoints using
ten labeled traces per website. Supervised baselines are trained on
the same support sets, and all methods are tested on the remaining
traces.

Cross-dataset transfer entails broader distribution shifts, with
simultaneous changes in website sets, collection periods, and network
paths between source and target data. Under these combined shifts,
existing pretrained attacks show limited transferability: their
highest average macro-F1 is only 53.6\%
(Table~\ref{tab:closed-world-transfer}).

The window-based methods RF and CountMamba are the strongest supervised
baselines, achieving average macro-F1 scores of 67.5\% and 63.4\%,
respectively. Both outperform supervised attacks that use packet or
burst sequences as inputs, suggesting that window-level traffic
aggregation supports more effective few-shot classification across
heterogeneous datasets. By comparison, Swallow and TraVerse, which
use ResNet-18 and an LLM backbone, respectively, achieve average
macro-F1 scores of only 44.5\%--52.4\% and 21.4\%--24.5\% across
the two pretraining sources. Their limited performance suggests that
image- and language-oriented backbones struggle to acquire transferable
traffic semantics even after traffic-domain pretraining.

In contrast, SCSM pretrained on GTT23 and Swallow D1 achieves the top
two accuracy and macro-F1 scores on every target dataset. The corresponding
average macro-F1 scores are 78.8\% and 81.3\%, exceeding RF by 11.3 and
13.8 percentage points, respectively. Built on WTCM representations
and a traffic-native state-space encoder, SCSM learns from diverse
pretraining views that vary global traffic organization and temporal
granularity while preserving the local traffic dynamics of real trace
fragments. Contrastive pretraining across these traffic-grounded views
promotes generalizable and transferable representations for few-shot
adaptation across datasets.

\subsubsection{Open-World Evaluation}
\begin{figure*}[!t]
  \centering
  \includegraphics[width=0.92\textwidth]{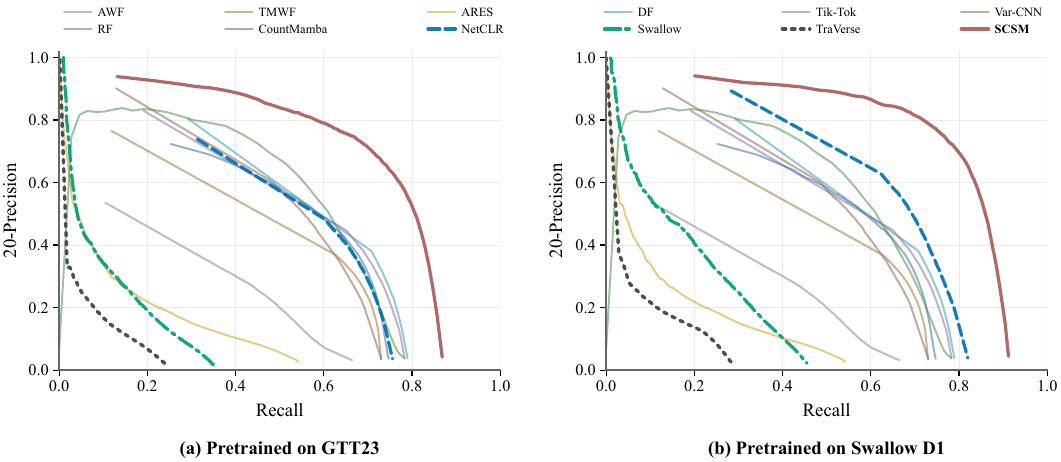}
  \caption{Open-world cross-domain transfer on Open-World DFSet with
  ten labeled traces per monitored website and an equal-sized
  unmonitored support set. The curves report $20$-precision versus
  recall for transferable attacks pretrained on (a) GTT23 and
  (b) Swallow D1. The same supervised baselines appear in both panels.}
  \label{fig:open-world-r-precision}
\end{figure*}

We evaluate whether pretrained representations transfer to open-world
recognition, which requires identifying monitored websites while
rejecting unmonitored traffic. On Open-World DFSet, adaptation uses
ten labeled traces per monitored website and an equal-sized set of
unmonitored support traces. Fig.~\ref{fig:open-world-r-precision}
compares the $20$-precision--recall curves for attacks pretrained on
GTT23 or Swallow D1, alongside supervised baselines, with $r=20$
as defined in \eqref{eq:r-precision}.

\begin{table*}[!t]
  \caption{Macro-F1 of WF Attacks under 10-Shot Transfer to Defended DFSet.}
  \label{tab:robust-defenses}
  \centering
  \scriptsize
  \setlength{\tabcolsep}{6.5pt}
  \renewcommand{\arraystretch}{1.08}
  \begin{tabular}{@{}ll*{9}{c}@{}}
    \toprule
    \multirow{2}{*}{Method} & \multirow{2}{*}{Source}
      & \multicolumn{3}{c}{Splitting Traffic Defenses}
      & \multicolumn{5}{c}{Disturbing Traffic Defenses}
      & \multirow{2}{*}{Average} \\
    \cmidrule(lr){3-5}\cmidrule(l){6-10}
    & & RR & BD & BWR & WTF-PAD & FRONT & RegulaTor
      & Tamaraw & Palette & \\
    \midrule
    AWF        & Target only & 0.549 & 0.124 & 0.032 & 0.211 & 0.098 & 0.060 & 0.066 & 0.061 & 0.150 \\
    TMWF       & Target only & 0.498 & 0.118 & 0.037 & 0.502 & 0.212 & 0.079 & 0.068 & 0.085 & 0.200 \\
    ARES       & Target only & 0.418 & 0.315 & 0.141 & 0.404 & 0.343 & 0.177 & 0.050 & \textbf{0.110} & 0.245 \\
    DF         & Target only & 0.540 & 0.125 & 0.038 & 0.506 & 0.239 & 0.077 & 0.069 & 0.081 & 0.209 \\
    Tik-Tok    & Target only & 0.616 & 0.434 & 0.066 & 0.525 & 0.221 & 0.104 & \underline{0.070} & 0.087 & 0.265 \\
    Var-CNN    & Target only & 0.121 & 0.219 & 0.011 & 0.247 & 0.115 & 0.087 & 0.014 & 0.057 & 0.109 \\
    RF         & Target only & 0.780 & 0.556 & 0.155 & 0.685 & \textbf{0.625} & \textbf{0.301} & 0.062 & 0.070 & \underline{0.404} \\
    CountMamba & Target only & 0.734 & 0.506 & 0.162 & 0.642 & 0.467 & 0.216 & \textbf{0.074} & 0.092 & 0.362 \\
    \midrule
    NetCLR   & GTT23      & 0.487 & 0.126 & 0.044 & 0.406 & 0.207 & 0.088 & 0.067 & 0.082 & 0.188 \\
    NetCLR   & Swallow D1 & 0.575 & 0.126 & 0.043 & 0.489 & 0.245 & 0.092 & 0.069 & 0.081 & 0.215 \\
    Swallow  & GTT23      & 0.382 & 0.182 & 0.033 & 0.322 & 0.264 & 0.147 & 0.061 & 0.068 & 0.183 \\
    Swallow  & Swallow D1 & 0.608 & 0.285 & 0.068 & 0.463 & 0.421 & 0.072 & 0.057 & 0.070 & 0.256 \\
    TraVerse & GTT23      & 0.161 & 0.088 & 0.012 & 0.016 & 0.013 & 0.009 & 0.001 & 0.026 & 0.041 \\
    TraVerse & Swallow D1 & 0.176 & 0.077 & 0.015 & 0.015 & 0.014 & 0.010 & 0.001 & 0.025 & 0.042 \\
    \midrule
    \textbf{SCSM} & GTT23
      & \underline{0.793} & \underline{0.562} & \underline{0.167}
      & \underline{0.697} & 0.557 & \underline{0.278} & 0.065
      & \underline{0.107} & 0.403 \\
    \textbf{SCSM} & Swallow D1
      & \textbf{0.812} & \textbf{0.596} & \textbf{0.170}
      & \textbf{0.774} & \underline{0.607} & 0.270 & 0.069 & 0.103
      & \textbf{0.425} \\
    \bottomrule
  \end{tabular}
\end{table*}

The source comparison highlights the role of pretraining data quality.
NetCLR improves markedly with Swallow D1 pretraining, outperforming
the other baselines at moderate-to-high recall while remaining below
SCSM. Swallow D1 provides higher-quality pretraining data through
repeated website visits collected under fixed
browser and guard-relay settings~\cite{Swallow}, whereas GTT23
contains heterogeneous real-user traces, many of which are short and
may capture only part of a page load~\cite{GTT23}.

SCSM performs strongly with both pretraining sources. With GTT23
pretraining, its maximum recall at a $20$-precision of at least $80\%$
reaches $58.0\%$, compared with $29.2\%$ for DF and $27.5\%$ for RF.
SCSM thus correctly identifies substantially more monitored visits
under the same precision requirement, even when pretrained on
heterogeneous real-user traffic. With Swallow D1 pretraining, its
maximum recall increases to $71.6\%$ under the same requirement,
a gain of $13.6$ percentage points, while retaining its lead over
NetCLR and the other baselines.

These results show that SCSM learns transferable representations from
heterogeneous real-user traffic and benefits further from higher-quality
pretraining data. Segmentation and Combination synthesize diverse,
high-quality pretraining views from lower-quality source traces by
reorganizing and combining observed traffic fragments while preserving
their local packet dynamics. These views enrich the traffic patterns
available for contrastive learning, enabling the adapted model to
identify monitored websites and reject unmonitored traffic with
limited labeled supervision.

\begin{table*}[!t]
  \caption{P@$k$ and MAP@$k$ (\%) of WF Attacks in the Multi-Tab Browsing
  Environment for Closed-World and Open-World Scenarios.}
  \label{tab:multi-tab-transfer}
  \centering
  \scriptsize
  \setlength{\tabcolsep}{6.5pt}
  \renewcommand{\arraystretch}{1.02}
  \begin{tabular}{@{}lll*{4}{cc}@{}}
    \toprule
    \multirow{2}{*}{Scenario} & \multirow{2}{*}{Method}
      & \multirow{2}{*}{Source} & \multicolumn{2}{c}{2-tab}
      & \multicolumn{2}{c}{3-tab} & \multicolumn{2}{c}{4-tab}
      & \multicolumn{2}{c}{5-tab} \\
    \cmidrule(lr){4-5}\cmidrule(lr){6-7}
    \cmidrule(lr){8-9}\cmidrule(l){10-11}
    & & & P@2 & MAP@2 & P@3 & MAP@3 & P@4 & MAP@4 & P@5 & MAP@5 \\
    \midrule
    \multirow{16}{*}{Closed-world} & AWF & Target only
      & 7.23 & 7.79 & 6.52 & 6.96 & 6.93 & 7.41 & 7.78 & 8.37 \\
    & TMWF & Target only
      & 31.36 & 36.94 & 21.33 & 25.71 & 18.65 & 22.70 & 15.24 & 18.33 \\
    & ARES & Target only
      & 19.86 & 22.91 & 5.59 & 5.61 & 9.25 & 10.15 & 8.99 & 9.51 \\
    & DF & Target only
      & 32.70 & 39.40 & 20.26 & 24.87 & 15.01 & 17.93 & 12.95 & 15.25 \\
    & Tik-Tok & Target only
      & 35.21 & 42.13 & 21.51 & 26.52 & 17.92 & 21.74 & 15.61 & 18.92 \\
    & Var-CNN & Target only
      & 25.39 & 30.54 & 18.58 & 23.24 & 14.35 & 17.32 & 11.39 & 13.20 \\
    & RF & Target only
      & 2.01 & 2.01 & 2.79 & 2.86 & 3.84 & 3.75 & 4.87 & 4.84 \\
    & CountMamba & Target only
      & 34.76 & 42.23 & 17.74 & 21.52 & 15.21 & 18.25 & 11.37 & 13.09 \\
    \cmidrule(l){2-11}
    & NetCLR & GTT23
      & 25.08 & 29.44 & 17.64 & 21.65 & 14.97 & 17.96 & 13.17 & 15.58 \\
    & NetCLR & Swallow D1
      & 27.85 & 33.13 & 18.80 & 23.10 & 15.61 & 18.85 & 13.45 & 15.89 \\
    & Swallow & GTT23
      & 19.11 & 22.81 & 14.40 & 17.78 & 12.52 & 15.19 & 10.22 & 11.58 \\
    & Swallow & Swallow D1
      & 26.90 & 32.84 & 16.39 & 20.55 & 13.60 & 16.59 & 10.84 & 12.59 \\
    & TraVerse & GTT23
      & 2.46 & 2.48 & 3.44 & 3.51 & 4.70 & 4.80 & 5.41 & 5.32 \\
    & TraVerse & Swallow D1
      & 2.47 & 2.45 & 3.45 & 3.52 & 4.75 & 4.83 & 5.53 & 5.57 \\
    \cmidrule(l){2-11}
    & \textbf{SCSM} & GTT23
      & \underline{49.07} & \underline{59.77} & \underline{34.62} & \underline{45.06} & \underline{29.18} & \underline{38.85} & \underline{22.98} & \underline{30.04} \\
    & \textbf{SCSM} & Swallow D1
      & \textbf{53.47} & \textbf{64.39} & \textbf{37.34} & \textbf{48.71} & \textbf{30.00} & \textbf{39.77} & \textbf{23.67} & \textbf{31.05} \\
    \midrule
    \multirow{16}{*}{Open-world} & AWF & Target only
      & 7.98 & 8.50 & 6.99 & 7.63 & 7.17 & 7.66 & 7.78 & 8.33 \\
    & TMWF & Target only
      & 28.49 & 33.09 & 20.70 & 25.18 & 17.03 & 20.53 & 15.15 & 18.29 \\
    & ARES & Target only
      & 17.53 & 20.17 & 9.87 & 11.59 & 11.03 & 12.69 & 10.47 & 11.99 \\
    & DF & Target only
      & 31.07 & 37.13 & 19.40 & 24.04 & 16.08 & 19.47 & 12.59 & 14.74 \\
    & Tik-Tok & Target only
      & 33.15 & 39.07 & 21.06 & 26.13 & 17.36 & 21.37 & 13.89 & 16.47 \\
    & Var-CNN & Target only
      & 22.98 & 27.19 & 16.28 & 19.96 & 14.19 & 17.18 & 10.87 & 12.71 \\
    & RF & Target only
      & 5.65 & 7.51 & 5.06 & 6.86 & 5.17 & 6.62 & 5.67 & 6.84 \\
    & CountMamba & Target only
      & 34.01 & 41.15 & 22.00 & 27.52 & 16.00 & 19.29 & 11.90 & 13.68 \\
    \cmidrule(l){2-11}
    & NetCLR & GTT23
      & 25.34 & 29.85 & 16.67 & 20.24 & 14.76 & 17.82 & 13.32 & 15.86 \\
    & NetCLR & Swallow D1
      & 27.03 & 31.93 & 17.84 & 21.91 & 15.38 & 18.58 & 13.38 & 15.87 \\
    & Swallow & GTT23
      & 24.10 & 29.28 & 13.76 & 16.85 & 12.14 & 14.46 & 9.92 & 11.24 \\
    & Swallow & Swallow D1
      & 24.37 & 29.28 & 15.53 & 19.26 & 13.32 & 16.16 & 11.12 & 12.80 \\
    & TraVerse & GTT23
      & 5.60 & 7.15 & 5.04 & 6.17 & 5.54 & 6.47 & 6.08 & 6.68 \\
    & TraVerse & Swallow D1
      & 5.55 & 7.12 & 5.02 & 6.18 & 5.68 & 6.66 & 6.07 & 6.75 \\
    \cmidrule(l){2-11}
    & \textbf{SCSM} & GTT23
      & \underline{47.14} & \underline{57.01} & \underline{32.84} & \underline{42.82} & \textbf{29.41} & \textbf{39.18} & \underline{23.42} & \underline{30.91} \\
    & \textbf{SCSM} & Swallow D1
      & \textbf{51.33} & \textbf{61.58} & \textbf{35.13} & \textbf{45.68} & \underline{29.34} & \underline{39.01} & \textbf{24.91} & \textbf{32.77} \\
    \bottomrule
  \end{tabular}
\end{table*}

\subsubsection{Robust Attack Evaluation}
We evaluate robustness to WF defenses by transferring models
pretrained on undefended GTT23 or Swallow D1 traffic to Defended
DFSet. For each defense, the pretrained models are independently
adapted using ten labeled traces per website. We consider six defenses, including
TrafficSliver~\cite{trafficsliver} with three splitting strategies:
Round Robin (RR), By Direction (BD), and Batched Weighted Random
(BWR). Table~\ref{tab:robust-defenses} reports macro-F1 across the
resulting eight configurations.

The existing pretrained baselines achieve an average macro-F1 of
at most $25.6\%$, indicating limited transferability to defended
traffic. SCSM achieves $40.3\%$ and $42.5\%$ when pretrained on
GTT23 and Swallow D1, respectively. The latter yields absolute
macro-F1 gains of $2.1\%$ over the strongest supervised baseline,
RF, and $16.9\%$ over the strongest transferable baseline, Swallow.

Under traffic splitting, the two pretrained SCSM models achieve
the highest and second-highest macro-F1 scores across all three
strategies. The Swallow D1-pretrained model reaches $81.2\%$,
$59.6\%$, and $17.0\%$ under RR, BD, and BWR, respectively,
exceeding RF by $3.2\%$, $4.0\%$, and $1.5\%$. Segmentation
allows the encoder to learn from observed traffic fragments, while
Masking encourages it to integrate evidence across partially visible
views. Together, these mechanisms support adaptation to the
incomplete traces observable on a single circuit.

Under traffic perturbation, RF and CountMamba generally outperform
sequence-based supervised attacks, suggesting that window-level
aggregation provides a more robust basis for few-shot recognition
by reducing dependence on exact packet arrangements. SCSM combines
windowed representations with traffic-grounded pretraining and
multi-scale adaptation. Its Swallow D1-pretrained model achieves
a macro-F1 of $77.4\%$ under WTF-PAD, exceeding RF by $8.9\%$,
and also outperforms the other pretrained attacks under FRONT and
RegulaTor. All methods obtain low macro-F1 scores under Tamaraw
and Palette, which suppress website-specific patterns through
padding and traffic regularization at the cost of bandwidth and
latency overhead~\cite{tamaraw,Palette}.

SCSM ranks first in four of the eight configurations. These results
demonstrate that representations learned from undefended traffic
support few-shot adaptation to both traffic splitting and
perturbation.

\begin{table*}[!t]
  \caption{Ablation Study of SCSM (\%).
  (a) Multi-level view construction.
  (b) Pretraining and few-shot adaptation.}
  \label{tab:ablation-study}
  \label{tab:ablation-view-construction}
  \label{tab:ablation-pretraining-scaling}
  \centering
  \footnotesize
  \begin{minipage}[t]{0.48\textwidth}
  \centering
  (a) Multi-level view construction\par
  \begin{tabular*}{\linewidth}[t]{@{\extracolsep{\fill}}lcc@{}}
    \toprule
    \multirow{2}{*}{Configuration}
      & \multicolumn{2}{c}{Macro centroid accuracy (\%)} \\
    \cmidrule(lr){2-3}
    & GTT23 & Swallow \\
    \midrule
    NetCLR & 20.29 & 65.63 \\
    Swallow & 35.74 & 60.14 \\
    TraVerse & 42.66 & 59.86 \\
    \midrule
    SCSM w/o augmentation & 22.14 & 5.16 \\
    SCSM w/ Masking only & 35.47 & 16.68 \\
    SCSM w/o Segmentation & 55.02 & 67.28 \\
    SCSM w/o Combination & 50.31 & 62.15 \\
    SCSM w/o Scaling & 51.45 & 52.98 \\
    SCSM w/o Masking & \textbf{59.02} & \textbf{69.83} \\
    \midrule
    SCSM & 53.83 & 69.03 \\
    \bottomrule
  \end{tabular*}
  \end{minipage}\hfill
  \begin{minipage}[t]{0.48\textwidth}
  \centering
  (b) Pretraining and few-shot adaptation\par
  \setlength{\tabcolsep}{3pt}
  \renewcommand{\arraystretch}{1.885}
  \begin{tabular*}{\linewidth}[t]{@{\extracolsep{\fill}}ccccc@{}}
    \toprule
    Pretraining & Fine-tuning & Inference & GTT23 & Swallow \\
    \cmidrule(lr){1-1}\cmidrule(lr){2-2}\cmidrule(lr){3-3}
    \multirow{2}{*}{\shortstack{Contrastive\\learning}}
    & \multirow{2}{*}{\shortstack{Temporal\\Scaling}}
    & \multirow{2}{*}{\shortstack{Multi-scale\\ensemble}}
    & Accuracy@3 & Macro-F1 \\
    & & & (\%) & (\%) \\
    \midrule
    \cmark & \cmark & \cmark & \textbf{79.25} & \textbf{92.21} \\
    \xmark & \cmark & \cmark & 65.08 & 88.82 \\
    \cmark & \xmark & \xmark & 75.97 & 85.36 \\
    \cmark & \cmark & \xmark & 76.95 & 91.29 \\
    \bottomrule
  \end{tabular*}
  \end{minipage}
\end{table*}

\begin{figure*}[!t]
  \centering
  \includegraphics[width=\textwidth]{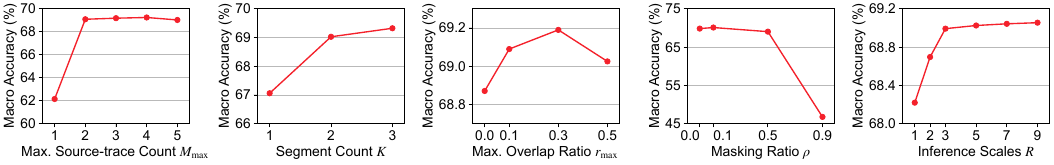}
  \caption{Sensitivity of pretrained SCSM representations to
  view-construction parameters and inference scales on Swallow.
  Results show mean macro centroid classification accuracy over
  six evaluation weeks.}
  \label{fig:parameter-sensitivity}
\end{figure*}

\subsubsection{Multi-Tab Attack Evaluation}
We evaluate transfer from single-tab pretraining to multi-tab identification
on ARESSet under closed-world and open-world scenarios with 2--5 tabs.
Table~\ref{tab:multi-tab-transfer} reports P@$k$ and MAP@$k$, with $k$
equal to the number of tabs.

Existing pretrained baselines underperform the best supervised attack on
both metrics in every setting. For example, in the closed-world 2-tab
scenario, NetCLR pretrained on Swallow D1 achieves the highest P@2 among
these baselines at 27.85\%, below Tik-Tok's 35.21\%. This gap suggests
that pretraining on individual single-tab traces provides limited support
for identifying multiple websites from mixed traffic. When traffic from
different tabs overlaps, the temporal patterns of individual page loads
are disrupted, limiting the transferability of representations learned
from isolated traces.

In contrast, the two SCSM models pretrained on GTT23 and Swallow D1
achieve the top two scores for both metrics in every setting. In the
closed-world 2-tab scenario, the Swallow D1-pretrained model achieves
53.47\% P@2 and 64.39\% MAP@2, with absolute gains of 18.26\% and
22.16\% over the best supervised score for each metric, respectively.
In the open-world counterpart, it reaches 51.33\% P@2 and 61.58\% MAP@2,
with corresponding absolute gains of 17.32\% and 20.43\%. SCSM retains
its lead as the number of tabs increases: in the open-world 5-tab setting,
its MAP@5 reaches 32.77\%, an absolute gain of 14.48\% over the best
supervised result of 18.29\%.

By combining multiple single-tab traces, Combination exposes SCSM's encoder
to mixed traffic during pretraining. Contrastive learning aligns
representations of views that share the same website composition but differ
in temporal arrangement, encouraging the encoder to capture website
information despite variations in traffic overlap. This supports transfer
from single-tab source data to multi-tab identification.

\subsection{Ablation Study}
\label{sec:ablation-study}
Table~\ref{tab:ablation-study}(a) assesses website separability using
mean macro centroid classification accuracy, with centroids computed from
pretraining data and equal weighting across websites and six evaluation
weeks. Table~\ref{tab:ablation-study}(b) reports mean 10-shot Accuracy@3
on GTT23 and macro-F1 on Swallow over six target weeks and scenarios,
respectively.

\subsubsection{Multi-Level View Construction}
The full pipeline outperforms the unaugmented baseline, with Combination
and Scaling benefiting both datasets. Segmentation improves Swallow
accuracy but lowers it on GTT23, where partitioning and reordering short,
incomplete traces may weaken the available website evidence.

Masking alone improves over the unaugmented baseline, yet removing it
from the full pipeline at the default $\rho=0.5$ raises accuracy on both
datasets. Excessive Masking may reduce the website evidence shared
between constructed views. On Swallow, $\rho=0.1$ outperforms both the
default and no Masking (Fig.~\ref{fig:parameter-sensitivity}), suggesting
a benefit from lighter Masking in the combined pipeline.

\subsubsection{Pretraining and Few-Shot Adaptation}
On both datasets, pretraining provides a better initialization for
few-shot adaptation than training from scratch. Temporal Scaling during
fine-tuning further improves performance by exposing the model to varying
temporal resolutions, while averaging logits across scales at inference
provides additional gains.

\subsection{Parameter Sensitivity}
\label{sec:parameter-sensitivity}
Fig.~\ref{fig:parameter-sensitivity} evaluates the sensitivity of pretrained
SCSM representations to five parameters on Swallow, using mean macro
centroid classification accuracy over six evaluation weeks.

For view construction, combining multiple traces and increasing $K$
from one to two improve accuracy, but larger $M_{\max}$ or $K$ provide
little further benefit. Accuracy also changes little across the tested
overlap bounds, suggesting limited returns from greater structural
variation. Masking intensity has a stronger effect: $\rho=0.1$ yields
the highest observed mean, whereas $\rho=0.9$ sharply reduces accuracy.
Excessive Masking may therefore remove website evidence needed for
discriminative representations.

At inference, accuracy also improves with $R$, but gains diminish beyond
five scales. The default $R=5$ therefore captures most of the observed
benefit without requiring a larger ensemble.

\section{Conclusion}
\label{sec:conclusion}
We presented SCSM, a traffic-native foundation model for transferable
website fingerprinting across changing deployment conditions. SCSM combines
multi-level view construction with contrastive pretraining of a state-space
encoder on windowed traffic counting matrices. Its views preserve the local
packet dynamics of observed trace fragments while varying their global
organization, temporal granularity, and visibility, enabling the encoder
to learn reusable representations from unlabeled traffic. Few-shot
fine-tuning adapts these representations to downstream tasks, while
multi-scale inference combines evidence across temporal resolutions.
Extensive experiments demonstrate SCSM's advantage over existing attacks
under temporal drift and across scenarios and domains, including open-world
recognition, defended traffic, and multi-tab browsing. These findings
highlight traffic-grounded self-supervised pretraining as a promising
approach to website fingerprinting across heterogeneous environments
with limited target-domain supervision.

\section*{Acknowledgments}
This work is supported by SJTU-QI'ANXIN Joint Lab of Information System
Security.

\bibliographystyle{IEEEtran}
\bibliography{references}

\end{document}